\documentstyle[12pt,epsf]{article}

\newcommand{\doublespace}{
  \renewcommand{\baselinestretch}{1.6}\large\normalsize}
 
\begin{document}

\def\be		{\begin{equation}}
\def\ee		{\end{equation}}
\newlength{\figsize}
\figsize = 0.75\textwidth

\begin{titlepage}

\begin{tabbing}
\` {\sl hep-lat/9709009} \\
    \\
\` LSUHE No. 262-1997 \\
\` OUTP-97-42P \\
\` September, 1997 \\
\end{tabbing}
 
\vspace*{1.0in}
 
\begin{center}
{\bf Magnetic monopole clusters, and monopole dominance after smoothing
in the maximally Abelian gauge of SU(2).\\}
\vspace*{.5in}
\underline{A. Hart}$^1$ and M. Teper$^2$\\
\vspace*{.2in}
$^1${\it Department of Physics and Astronomy, Louisiana State University,\\
Baton Rouge. LA 70803 U.S.A.\\
email: hart@rouge.phys.lsu.edu\\}
\vspace*{.2in}
$^2${\it Theoretical Physics, University of Oxford,\\
1 Keble Road, Oxford. OX1 3NP U.K.\\
email: teper@thphys.ox.ac.uk\\}
\vspace*{0.2in}
{\it (To be published in the proceedings of Lattice 97 --- 
Nucl. Phys. B (Proc. Suppl.).}
\end{center}

\hspace{2.0in}
\end{titlepage}
\vfill\eject

\doublespace
\pagestyle{empty}

\begin{center}
{\bf Abstract}
\end{center}

In the maximally Abelian gauge of SU(2), the clusters of monopole
current are found to divide into two distinct classes. The largest
cluster permeates the lattice, has a density that scales and produces
the string tension. The remaining clusters possess an approximate
$1/l^3$ number density distribution ($l$ is the cluster length), their
radii vary as $\surd l$ and their total current density does not
scale. Their contribution to the string tension is compatible with
being exactly zero. Their number density can be thought of as arising
from an underlying scale invariant distribution. This suggests that
they are not related to instantons.  We also observe that when we
locally smoothen the SU(2) fields by cooling, the string tension due
to monopoles becomes much smaller than the SU(2) string tension. This
dramatic loss of Abelian/monopole dominance occurs even after just one
cooling step.

\vspace*{.5in}
 
\setcounter{page}{0}
\newpage
\pagestyle{plain}


Magnetic monopole currents
\cite{degrand80}
are central to the dual superconducter hypothesis for confinement
--- as seen after Abelian projection
\cite{mandelstam76,thooft81}
to the maximally Abelian gauge (MAG)
\cite{kronfeld87a}.
Using the unambiguous division of the current into mutually
disconnected clusters, we find that only the single largest cluster
contributes to the string tension, $\sigma$.
We also find that the density of this largest cluster scales well 
and that the scaling violations seen in the total current density
are due to the smaller clusters (see also 
\cite{poulis97})
which do not contribute to $\sigma$ (see also
\cite{stack94}).

In each configuration we find that one cluster is significantly larger
than the rest (on an $L = 16$ lattice at $\beta = 2.3$ it has mean 
length 10169, compared to a mean of only 67 for the second largest). 
This large cluster fills the entire lattice volume and its 
length, at fixed $\beta$, is proportional to the volume, $L^4$.
By contrast the length of the second largest cluster increases
much more weakly with $L$ and it is much more localised.

\begin{figure}
	\begin{center}

	\leavevmode
	\epsfxsize = \figsize
	\epsffile{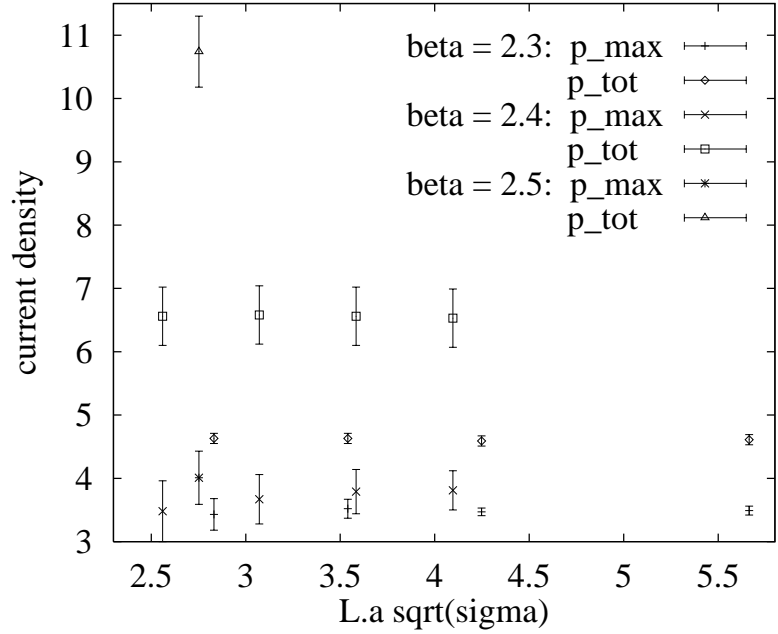}

	\end{center}

	\caption{Scaling with the lattice size of the total 
	current density and that of the largest cluster.}
	\label{fig_lmax_scaling}

	\end{figure}

If the total `length' of magnetic current is $l_{\hbox{\tiny tot}}$,
we may use the $SU(2)$ string tension to write the
current density in nonperturbatively defined physical units
\be
\rho_{\hbox{\tiny tot}}= 
{l_{\hbox{\tiny tot}}.a \sqrt \sigma \over (L.a \sqrt \sigma)^4} = 
{l_{\hbox{\tiny tot}} \over L^4 (a \sqrt \sigma)^3}.
\ee
Similarly we define $\rho_{\hbox{\tiny max}}$ for the largest cluster
alone. Whilst the total current density shows strong scaling
violations in Fig.~\ref{fig_lmax_scaling}, that of the largest cluster
alone shows remarkably good scaling from $\beta = 2.3$ to $2.5$.

The time--like monopole current links in a given time slice of the lattice
may be treated as a 3d gas of magnetic charges. Moving out from a
given charge, we observe charge screening characteristic of a
plasma. The screening length for the magnetic current from the largest
cluster alone is many times greater than that for the combined smaller
clusters.
Since the screening length is related to the string tension, this and
the scaling behaviour already suggest that the largest cluster plays
the dominant r\^ole in the infrared physics.

The separate contributions to the string tension from the largest
cluster, and from the remainder are shown in
Table~\ref{tab_sigma_prune}. The string tension is consistent with
being entirely due to the
largest cluster; particularly at smaller values of $\beta$ where
the distinction in length between the largest and second largest
clusters is clearest and the volumes are largest in physical units.

\begin{figure}
	\begin{center}

	\leavevmode
	\epsfxsize = \figsize
	\epsffile{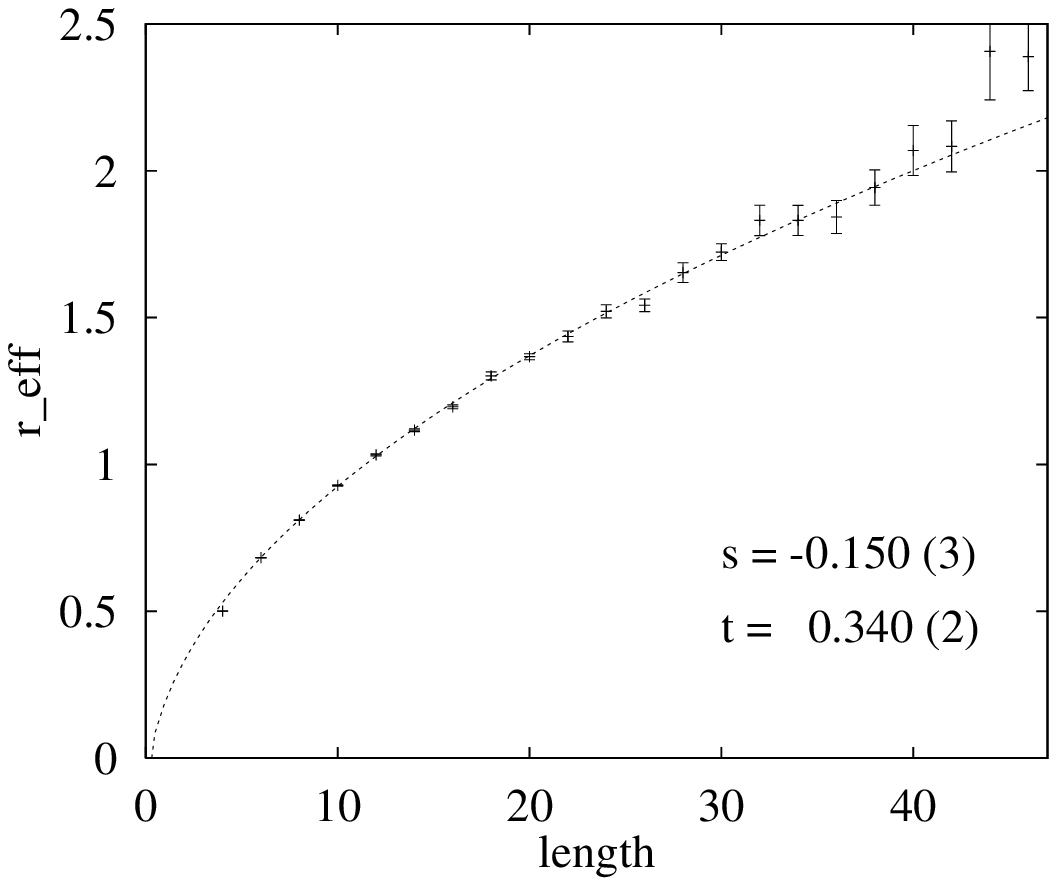}
	\end{center}

	\caption{The fit $r_{\hbox{\tiny eff}} = s + t \surd l$ 
	at $\beta = 2.3$ on $12^4$.}
	\label{fig_ext_2b3_l12}

	\end{figure}

We now calculate the effective radius, $r_{\hbox{\tiny eff}}$, for each
cluster, and plot its average value as a function of cluster length, $l$, in
Fig.~\ref{fig_ext_2b3_l12}. The data fits well a curve $r_{\hbox{\tiny
eff}} \sim \sqrt l$, with the parameters showing only a weak
dependence on $\beta$. This form is reminiscent of the displacement
of a particle undergoing a random walk of $l$ steps.

\begin{figure}
	\begin{center}

	\leavevmode
	\epsfxsize = \figsize
	\epsffile{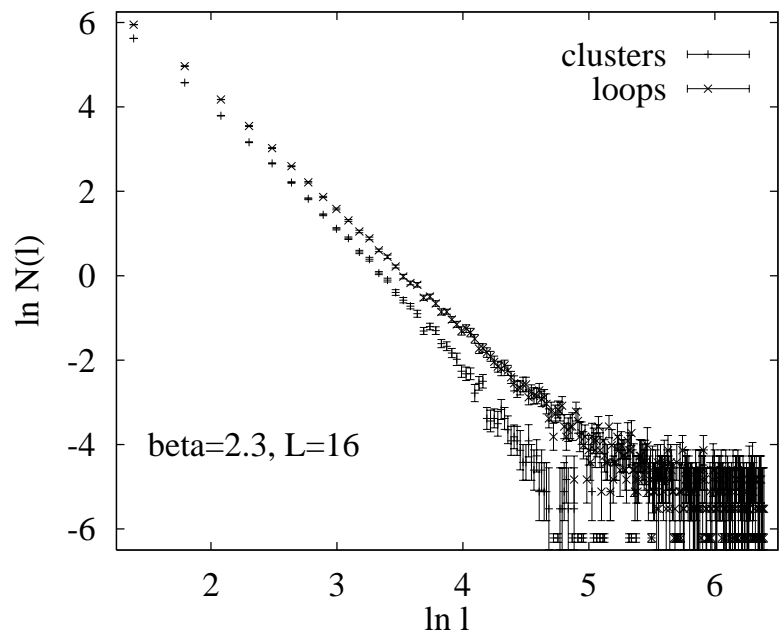}
	\end{center}

	\caption{Monopole loop and cluster spectra.}
	\label{fig_spectra_2b3_l16}

	\end{figure}

The cluster `spectrum', $n(l)$, is the mean
number of clusters of a given length on a configuration. It is
plotted versus $l$ in Fig.~\ref{fig_spectra_2b3_l16} and we see
that it is close to a power law
$n(l) = c/l^\gamma$
for values of $l$ where the errors are small
(similar to what was seen for current loops
\cite{hart97a}).
For all of our data $\gamma \in [2.85,3.15]$.

This exponent may be understood as arising from a general scale
invariant distribution of objects of radius $r$ in 4d. Using
$r_{\hbox{\tiny eff}} \sim \sqrt l$ then predicts $\gamma = 3$:
\be
n(r) dr \sim \frac{dr}{r} \times \frac{1}{r^4} 
\hbox{\hspace{1em}} \Rightarrow \hbox{\hspace{1em}}
n(l) dl \sim \frac{dl}{l^3}.
\ee
We recall that a semiclassical instanton of core size $\rho$ in 
the MAG generates a
monopole loop of proportional length within its core (e.g.
\cite{hart96}). 
More generally, we expect a cluster of radius $\propto \rho$.
A scale invariant distribution of core sizes might 
occur for large $\rho$ if the action freezes out there.
This does not explain, however, why it is the smaller 
clusters that most clearly fall on the power law.
Indeed the small instanton distribution is very far from being
scale invariant. Thus instantons are not an obvious source for
these monopole clusters.

%
%
%
%
%

We turn now to our second topic: the fate of monopole/Abelian dominance
on cooled or smoothened SU(2) fields. We recall that 
the interest in the MAG arises because here alone the SU(2) string
tension is nearly reproduced by the monopole currents
(e.g.
\cite{bali96}).
This would seem to imply a correlation between the currents and the
long range properties of the SU(2) vacuum. It is interesting, in light
of this, to investigate the behaviour of the SU(2) and monopole string
tensions under the application of small, local deformations of the
vacuum fields.

An ensemble of configurations was prepared using the Wilson action
${\cal S}$ and the SU(2) string tension was calculated. To each
configuration we applied a certain number of `smoothing'($\equiv$
`cooling') steps, to yield a new ensemble and the 
string tension was calculated afresh. Small amounts
of such smoothing merely remove the ultraviolet fluctuations in the
vacuum. The infrared properties, including the string tension and, by
definition, the lattice spacing, are unchanged.

Taking the smoothed ensemble, we fix to the MAG and obtain the
monopoles and their contribution to the string tension. To try to
minimise the (small) effects of Gribov copies, the procedure was
slightly more complicated. The original SU(2) ensemble was gauge fixed
and a single smoothing sweep applied. The configurations were then
gauge fixed again. (Since one smoothing step is a small perturbation,
it might be expected that the number of sweeps necessary to gauge fix
the second time would be smaller than the first but this did not seem to
be the case.) Additional smoothing proceeded accordingly.

To estimate the string tension, we here present initial results using
the effective string tension defined from the square Creutz ratios,
$\sigma_{\hbox{\tiny eff}}(r) = - \log {\hbox{Cr}}(r,r)$, which
approaches the asymptotic string tension for large $r$ (although
not necessarily from above on the smoothened fields where
`positivity' no longer holds).

\begin{figure}
	\begin{center}

	\leavevmode
	\epsfxsize = \figsize
	\epsffile{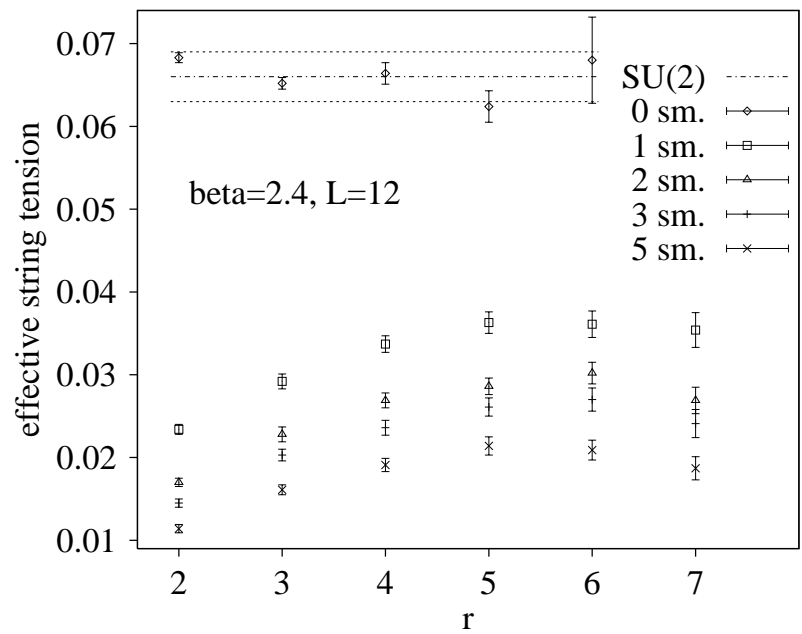}
	\end{center}

	\caption{Smoothing and the effective string tension.}
	\label{fig_eff_cool}

	\end{figure}

Fig.~\ref{fig_eff_cool} shows the monopole string tension after a
number of smoothing steps, and the unsmoothed SU(2) value. Even one
smoothing sweep causes a large reduction in the
monopole string tension. Further smoothing continues to reduce its
value, but at a decreased rate.

The infrared properties of the SU(2) ensemble have not changed (as
we see by an explicit calculation of 
the SU(2) string tension), yet the monopole physics is very
different. This may have implications for the interpretation of
monopole (and Abelian) dominance as a connexion between the long range
properties of the SU(2) vacuum and the magnetic currents.

These results agree with independent studies using
different techniques
\cite{kovacs97},
but contrast, however, with  
\cite{wensley96}
which used Metropolis cooling (a much less efficient way of
removing short distance fluctuations).  
Further investigation is clearly necessary here.

	\vskip 0.20in
	\noindent {\bf Acknowledgements}

	The work of A.G.H. was supported by the United States
	Department of Energy grant DE-FG05-91ER40617 and that of
	M.T. by United Kingdom PPARC grants GR/K55752 and GR/K95338.

\newpage

	\begin{table}
	\begin{center}
	\begin{tabular}
	{l*{2}{r@{.}l@{\hspace{0.5\tabcolsep}(}r@{)\hspace{2\tabcolsep}}}}
	\hline \hline
	$L=16$ & 
	\multicolumn{3}{c}{$\beta = 2.3$} &		
	\multicolumn{3}{c}{$\beta = 2.4$} \\
	\hline
	all: &		0&128 &  5 & 
			0&067 &  2 \\
	largest:  &	0&122 &  5 &
			0&058 &  2 \\
	rest: &		0&000 &  1 &
			0&001 &  1 \\
	\hline
	\% curr.: &	\multicolumn{3}{c}{75.7 (3)} & 
			\multicolumn{3}{c}{56.8 (3)} \\
	\hline \hline
	\end{tabular}
	\end{center}
	\caption{The monopole string tension, and the proportion 
		of the total current in the largest cluster.}
	\label{tab_sigma_prune}
	\end{table}

\end{document}